# Deformation characteristics and mechanical properties of a non-rigid square-twist origami structure with rotational symmetry

Shixi Zang[a,b,§], Jiayao Ma[a,b,§], Yan Chen[a,b,*]

[a] Key Laboratory of Mechanism Theory and Equipment Design of Ministry of Education, Tianjin University, 135 Yaguan Road, Tianjin, 300350, China

[b] School of Mechanical Engineering, Tianjin University, 135 Yaguan Road, Tianjin, 300350, China

[§] Joint first authors

**Abstract.** Non-rigid origami patterns could provide more versatile performance than their rigid counterparts in the design of mechanical metamaterials owing to the simultaneous deformation of facets and creases, but their complex deformation modes make quantitative characterization and programmability of mechanical properties a challenging task. Here, we investigated the tensile behavior of a non-rigid square-twist origami structure with rotational symmetry by combining biaxial tension experiments and finite element modeling. A three-stage deformation process, including tightening, unlocking, and flattening, of the structure was unveiled through a detailed analysis of facet distortion and crease rotation, and the relationship between structure deformation and several key features in the energy, force, and stiffness curve was obtained. Based on the analysis, an empirical model was built to correlate the geometric and material parameters of the structure and its deformation energy, initial peak force, and maximum stiffness, which were further validated through experiments. Using the model, the mechanical properties of the structure can be accurately predicted and programmed based on specific engineering requirements, thereby serving the development of new programmable mechanical metamaterials based on the family of square-twist origami.

---

[*] Corresponding author at: School of Mechanical Engineering, Tianjin University, Tianjin 300350, China
E-mail address: yan_chen@tju.edu.cn

# 1. Introduction

Mechanical metamaterials have enabled extraordinary and customizable properties beyond traditional materials owing to their special repeating microstructures independent of scale and base material [1-11]. By purposely engineering the kinematic motion, structural deformation, or equilibrium state transition of the microstructures, exotic properties and functionalities, including programmable reconfiguration [12], negative Poisson's ratio [13], compression-torsion conversion [14], structural vibration absorption [15], tunable buckling [16], can be achieved. To satisfy the increasingly demanding engineering requirements, the design of metamaterials has evolved from qualitatively obtaining a certain property to quantitatively controlling it by selecting geometric and material parameters.

Origami, owing to its superior capability of generating 3D structures from 2D sheets, has been widely studied as an important design strategy for metamaterials. Existing origami metamaterials are mainly developed from rigid origami patterns, such as the well-known Miura-ori pattern [17-20] and its derivative [21-24]. The rigid patterns can be folded by pure rotation of the creases without deformation of the facets. By treating the creases as revolute joints and facets as rigid links, the folding of a rigid pattern is essentially a mechanism motion, which can be quantified by various approaches such as numerical algorithms [25], quaternions and dual quaternions [26], matrix method [27], and kinematic method [28, 29]. When the rigid pattern has a single degree of freedom, its mechanical response can be accurately characterized based on the relationship among the rotation of the creases [13, 21, 22, 30]. Recently, non-rigid origami patterns, represented by the Kresling [31, 32] and square-twist patterns [33, 34], have been of increasing interest to researchers. In contrast to rigid patterns, the non-rigid ones are folded by simultaneous facet distortions and crease rotations, and therefore offer more versatile mechanical properties in a wider tunable range, including bistability [34, 35], graded stiffness [36], and high energy absorption [37]. However, the complicated deformation modes of non-rigid patterns make it a challenging task to develop analytical models for property characterization and programming. A commonly used approach is to triangulate the pattern by adding extra virtual creases in non-triangular facets [34]. This usually will turn a non-rigid pattern into a rigid one, but not necessarily with one degree of freedom. Then by assigning different rotational stiffness to the original and virtual creases, the deformation of the pattern can be solved analytically in certain cases [38]. The essence of this approach is using the rotation of virtual creases to simulate the bending of facets. Therefore, when the main deformation of the facets is bending with a single curvature, this approach will yield quite accurate results. However, when the facet distortion is more complex with non-zero Gauss curvature, or the non-rigid pattern is already formed by only triangular facets and cannot be further triangulated, such as the waterbomb pattern [39], it would be difficult to obtain realistic results out of this approach. When only one or a few facets in a non-rigid pattern are noticeably bent with a single curvature, it is possible to add virtual creases only to those facets and obtain an equivalent rigid pattern with a single degree of freedom. An elegant theoretical solution can be derived in this case [33]. Yet, there is no ready solution for every non-rigid pattern. As the complexity of the pattern increases, an analytical solution cannot always be obtained. In response to this difficulty, several computational approaches based on the same triangulation principle have been developed for non-rigid patterns, such as the bar-and-hinge model [40] or the pin-jointed bar framework model [41]. Those models enable an analysis of complicated patterns very efficiently in comparison with standard finite element models, but also inevitably preserve the limitation of pattern triangulation.

In this paper, we are focusing on a particular non-rigid origami pattern with four-fold rotational symmetry as shown in Fig. 1, also referred to as the type 1 square-twist pattern, which belongs to a family of four square-twist patterns with varying crease mountain-valley assignments and rigidity [42]. In addition to the applications in compliant mechanism [43], frequency reconfigurable origami antenna [44], and mechanical energy storage [45], this pattern has great potential in the design of mechanical metamaterials for two reasons. First of all, it has a strong self-locking behavior which is remarkably different from the other three members of the family. Moreover, it is relatively easy to combine it with the other rigid and non-rigid square-twist patterns to form metamaterials with a wide range of tunable properties. However, the quantitative relationship between the mechanical properties of the pattern and the geometric and material design parameters, which is essential for the programmability of the metamaterials, has so far been unclear. Previous efforts on mechanical characterization of the pattern include triangulating the entire pattern and analyzing it as a system of bars and hinges [34]. Alternatively, a virtual crease was introduced in the central square facet to turn it into a rigid pattern so that it can be analyzed using established mathematical [46] or kinematic approach [42]. Nevertheless, as will be shown later in the paper, those two approaches could not capture the complex facet deformation in the pattern. Finite element models of the pattern have also been developed for specific applications [43, 44], but little information is given about the detailed deformation process or validation with experiments. In view of this, we propose to characterize and program the mechanical properties of the pattern through a combination of experiments, finite element simulation, and empirical model development.

The outline of this paper is as follows. The biaxial tension experiment and numerical simulation of the type 1 origami structure are presented in Section 2. In Section 3 the deformation process of the structure, as well as the energy, force, and stiffness responses, are discussed in detail. In Section 4, an empirical model is built based on the experimental and numerical analysis to correlate the geometric and material parameters of the structure and its mechanical properties. Further validation of this model through experiments is also presented in this section. Finally, a conclusion is presented in Section 5 to end the paper.

## 2. Biaxial tension experiment and finite element modeling

The type 1 square-twist pattern in Fig. 1A consists of a central square facet, four trapezoidal ones, and four rectangular ones, which is parameterized by two side lengths, $l$ and $a$, and a twist angle, $\alpha$. A paper model in the deployed and folded configurations is shown in Fig. 1B. It can be seen that the crease mountain-valley assignment forms a four-fold rotational symmetry in the pattern. The dihedral angles of the facets at crease $i$ are described by $\varphi_i$ ($i$=1, 2, 3, …, 12), and the vertices are defined as $P_j$ ($j$=1, 2, 3, …, 16). To fully understand the deformation features and mechanical properties of the type 1 square-twist structure, both experiment and numerical simulation are conducted in this section.

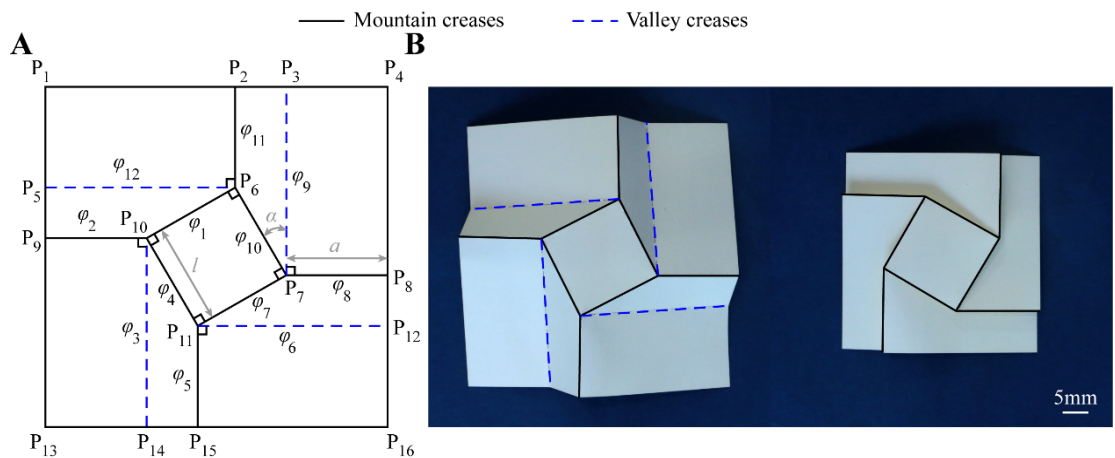

**Fig. 1.** (**A**) Pattern and crease mountain-valley assignment of the type 1 square-twist structure, (**B**) unfolded and folded configurations of a paper model with side lengths $l$=$a$=16.25mm and twist angle $\alpha$=30°.

## 2.1. Experimental setup

Considering the rotational symmetry of the pattern, a biaxial tension experiment, loading at four corners of the structure, was conducted through a specially designed loading mechanism. As shown in Fig. 2**A**, the square loading mechanism consisted of four sliding units, each of which was composed of a 3D-printed block and a steel linear guide fixed together. There was a channel in each block, where the linear guide from the neighboring sliding unit could pass through. The four corners of the specimen were respectively attached to the four blocks, and then the loading mechanism was connected to a horizontal testing machine by a stationary and a movable fixture. Six universal wheel bearings were installed at the bottom of the loading mechanism to minimize the friction between the surface of the machine and the loading mechanism. During loading, the blocks slid along the linear guides to stretch the specimen equally in the two diagonal directions, which is illustrated through the deformation process of the specimen in Fig. 2**B**. The horizontal test machine had a stroke of 800mm and a load cell of 300N with an accuracy of 0.5%. In the experiment, the specimen was tensioned by a displacement of $\Delta x_{max}$=23mm along the diagonal direction at a loading rate of 0.2mm/s. The exact deformations of the square facet, as well as the portions of the rectangular and trapezoidal ones that were not occluded by others, were captured by a digital image correlation (DIC) system CSI Vic-3D9M. The camera resolution of the DIC system was 2704×3384 pixels and the frame time interval was 500ms. Three specimens were tested to obtain reliable results.

The test specimen was manufactured by polyethylene terephthalate (PET) sheets of thickness $t$=0.4mm. The geometric parameters were selected as $l$=$a$=16.25mm and $\alpha$=30°. The creases were cut as 1.2-mm-wide dotted lines with 2mm perforations at 1.5mm intervals by a Trotec Speedy 300 laser cutter, and holes with 3.2mm in diameter were cut at the vertices to mitigate stress concentration and fracture. Afterward, the perforated sheet was folded by hand to the fully folded shape. Due to the elastic spring back of the creases, the specimen was not completely flat but had a natural dihedral angle formed by the square and trapezoidal facets $\varphi_{i0}$=19.58° ($i$=1, 4, 7, 10), diagonal length $L_{D0}$=53.30mm, and height $H_{D0}$=7.80mm.

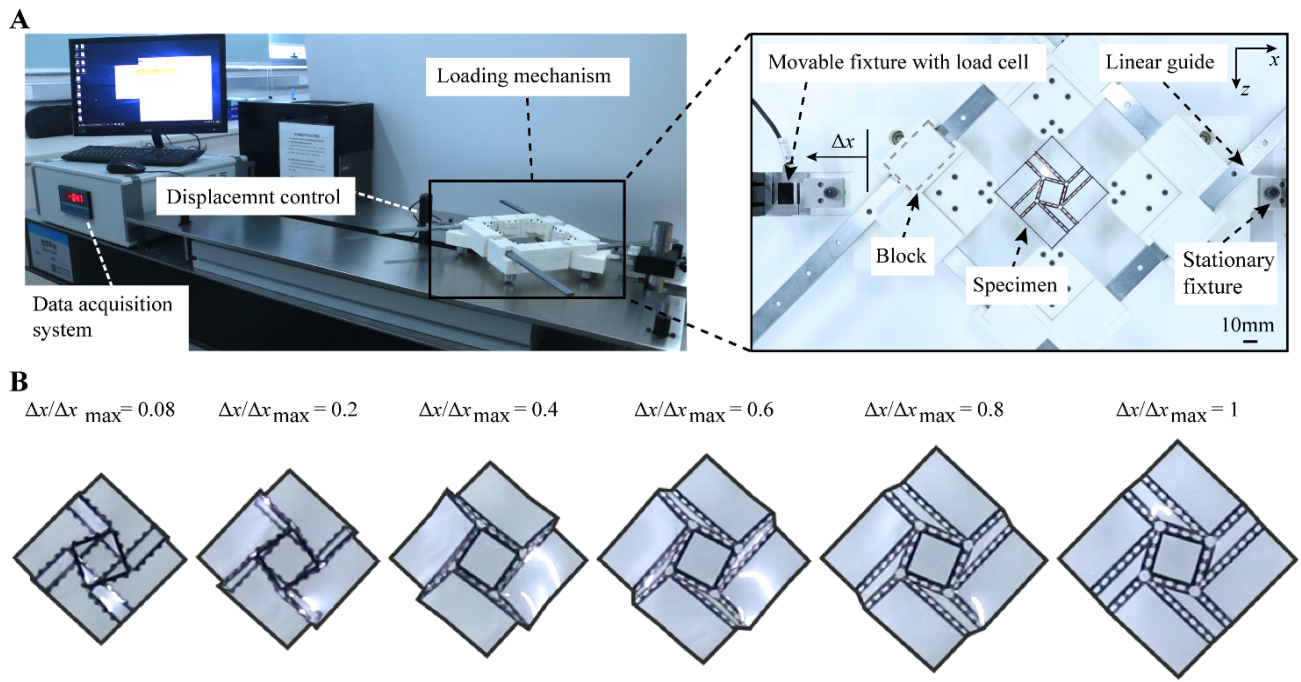

**Fig. 2.** (**A**) Experimental setup and details of the loading mechanism. (**B**) Deformation process of the specimen.

## 2.2 Finite element modeling

In addition to the experiment, a finite element model of the type 1 square-twist structure using Abaqus/Explicit was also developed, first of all, to obtain detailed deformations of the facet portions that were occluded by other facets during tensioning, and secondly, to investigate the effects of design parameters on the mechanical properties of the structure. As mentioned in reference [42], the non-rigid pattern satisfies the compatibility condition only at the fully folded and fully deployed configurations, where all the facets maintain the original flat shape. Then, for the type 1 specimen with a natural dihedral angle, a major difficulty arises, i.e., how to rationally build its geometry when the planar surface cannot be employed in all of the facets. Here the method we adopted was to keep the central square flat and use curved surfaces to replace the rectangular or trapezoid facets. To investigate the influence of curved facets on the mechanical properties of the structure, four different geometric construction schemes were designed, replacing the trapezoidal facets with two planar triangles as shown in Fig. 3**A**, replacing the rectangular facets with curved surfaces, replacing trapezoidal facets with curved surface, and replacing trapezoidal facets with curved creases and curved surfaces. The tension simulation results showed that the four models generated nearly identical force versus displacement curves, implying that the behavior of the structure was not sensitive to the specific type of curved surfaces in the geometric construction. Thus we adopted the scheme in Fig. 3**A** in the subsequent simulation. Details of the geometric construction methods and results are explained in Appendix A. The numerical model had an identical natural configuration and facet thickness with the physical specimen. The facets were modeled as thin shells with elastoplastic properties obtained from tensile tests: Young's modulus $E$=2216.78MPa and yield stress $\sigma_y$=24.46MPa. The Poisson's ratio, $v$, was set to be 0.39 [47]. The creases were modeled by revolute connection with tie constraint. The torsional stiffness per unit length of the creases, $k_c$=0.44N·rad$^{-1}$, was experimentally determined from a rigid type 3 square-twist structure that had identical geometry and base material with the type 1 specimen [33]. Multiple point constraint of a beam type as shown in Fig. 3**B** was applied to the four corners of the model to achieve a biaxial tension. Each loading point had only one translational degree of freedom in the *x-z* plane, i.e., the two points on the diagonal parallel to the *x*-axis moved along the *x*-direction, and the other two on the diagonal parallel to the *z*-axis moved along the *z*-direction. The four-node quadrilateral shell elements with reduced integration, S4R, were used to mesh the model. Mesh convergence was carried out to determine the element size of 1mm. And a loading rate of 50mm/s was adopted to ensure the loading process was quasi-static.

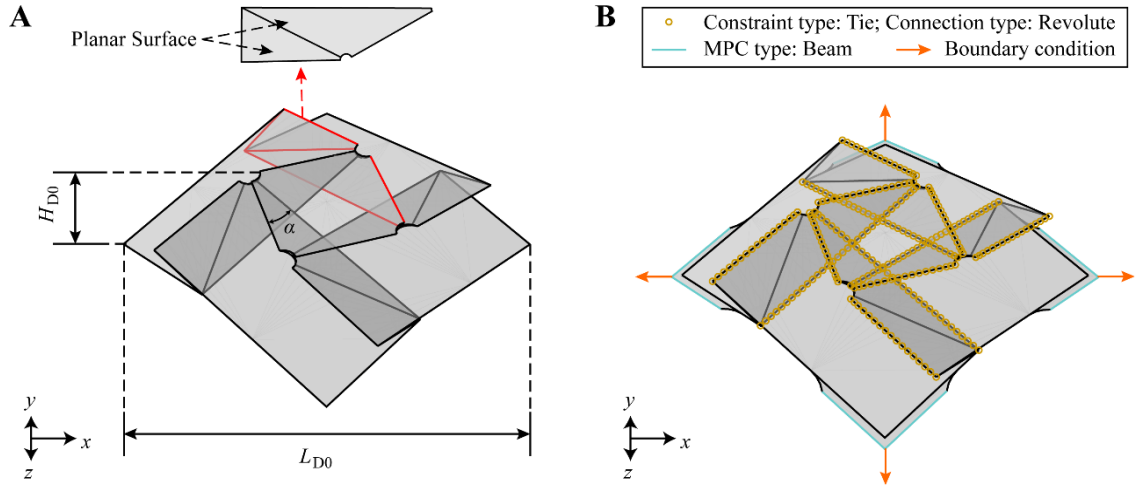

**Fig. 3.** (**A**) Geometric model of the type 1 square-twist structure constructed by replacing the trapezoidal facet with two intersected triangular planar surfaces. (**B**) Details of the constraint and boundary condition.

## 2.3. Validation of the finite element model

The experimentally reconstructed and numerically obtained square, rectangular, and trapezoidal facets are respectively compared in Fig. 4**A-C** at normalized displacement $\Delta x/\Delta x_{max}$=0.08, 0.2, 0.4, 0.6, and 0.8. The missing trapezoid data at $\Delta x/\Delta x_{max}$=0.08 and 0.8 is caused by the facet overlap and reflection of the specimen. A good match between the experimental and numerical results is observed in the entire unfolding process. The pie graphs below the configurations indicate that the fitting error is within half of the material thickness, $t$=0.4mm, in over 80% of the measured area. Therefore, the numerical model is capable of capturing the main deformation feature of the structure.

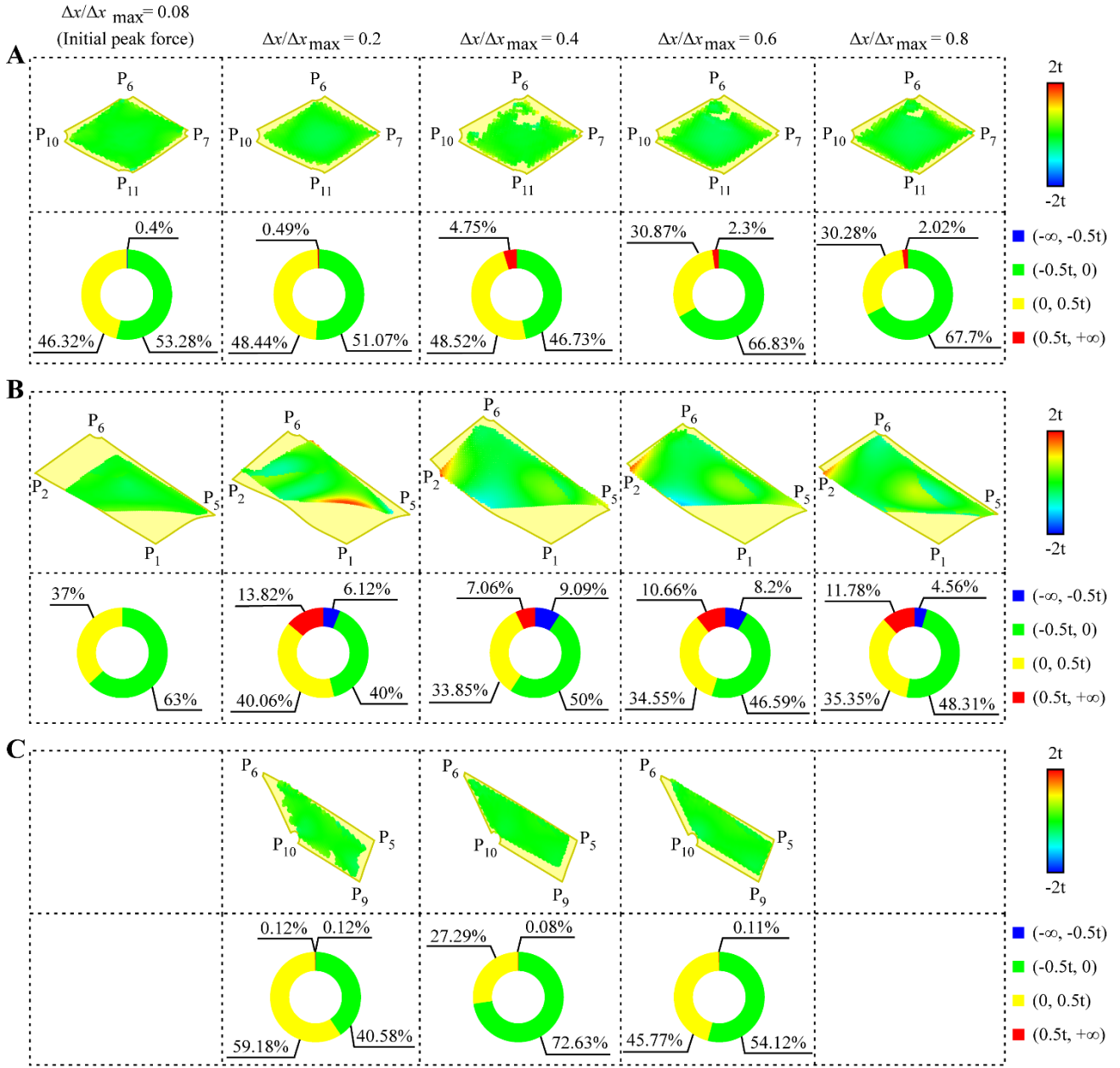

**Fig. 4.** Comparison between experimentally reconstructed and numerically obtained deformed shapes and errors (pie graphs) of the (**A**) square, (**B**) rectangular, and (**C**) trapezoidal facets.

Moreover, Figure 5**A**-**C** shows the numerical normalized energy, $U/(k_c l)$, normalized force, $F/k_c$, and normalized stiffness, $Kl/k_c$, drawn against the normalized displacement, $\Delta x/\Delta x_{max}$, of the structure together with the experimental ones. The force is directly measured from experiments or exported from the numerical model. Then, it is integrated and differentiated with respect to tension displacement to gain energy and stiffness, respectively. The red shades for the experimental curves represent the repeatability of three specimens. Again a reasonable agreement between numerical and experimental curves is obtained. Thus, it can be concluded that the numerical model can accurately capture the mechanical behaviors of the structure, which will later be used to unveil the detailed deformation process in the subsequent section.

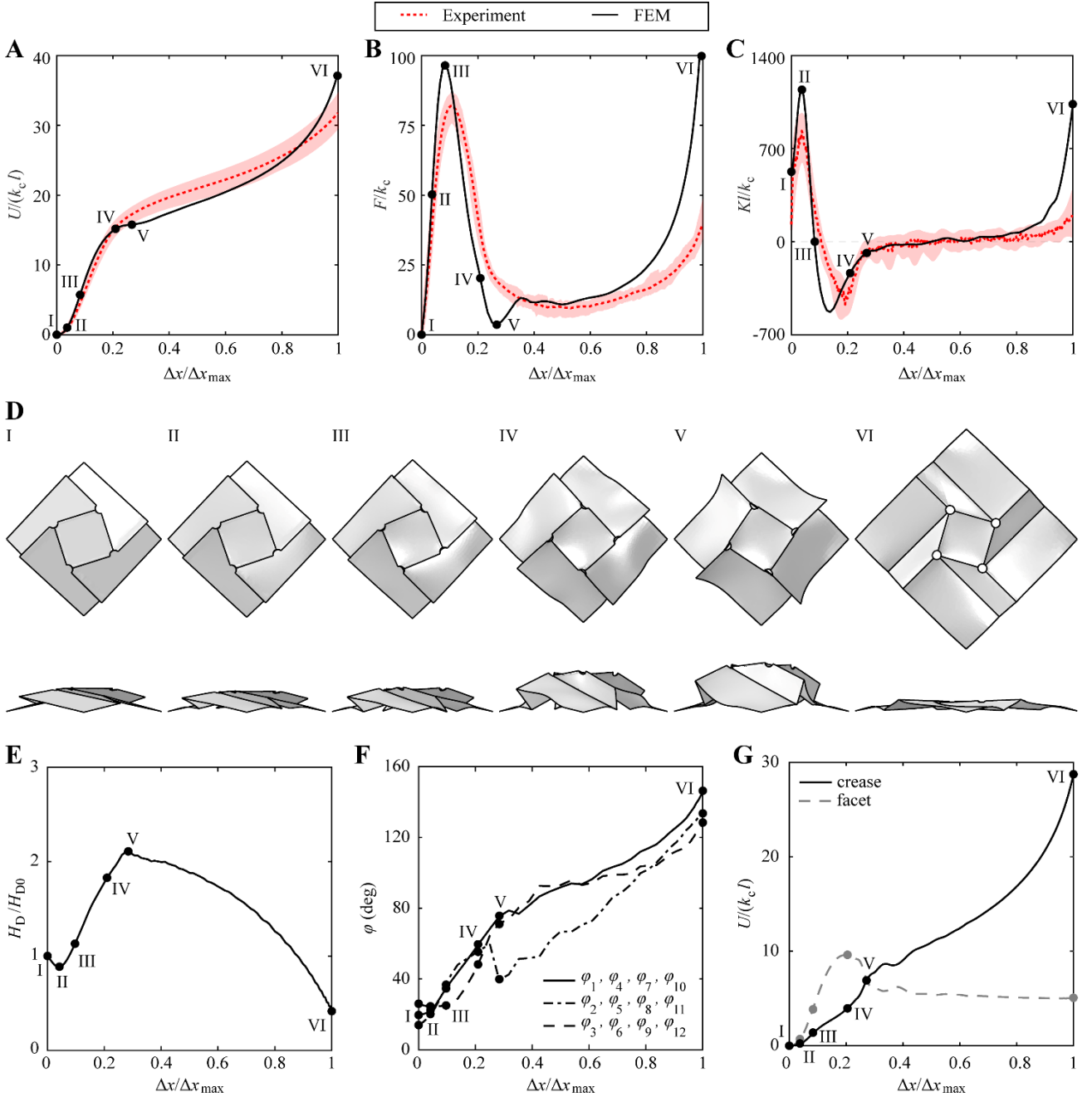

**Fig. 5.** (**A-C**) Experimental and numerical normalized energy, $U/(k_c l)$, normalized force, $F/k_c$, and normalized stiffness, $Kl/k_c$, of the type 1 square-twist structure against normalized displacement, $\Delta x/\Delta x_{max}$. (**D**) Six typical configurations of the structure in the front and side views. (**E-G**) Numerical normalized height, $H_D/H_{D0}$, dihedral angles, $\varphi_i$ ($i$=1, 2, …, 12), and facet and crease energies of the structure versus normalized displacement, $\Delta x/\Delta x_{max}$.

## 3. Deformation process

To demonstrate the deformation process of the origami structure, we select six key points marked with I to VI based on the numerical energy, force, and stiffness curves in Fig. 5**A-C**, where I and VI represent the initial and final configurations, II is the maximum stiffness, III and V are the initial peak and valley forces, and IV describes a transition point in the energy curve. The configurations of the structure corresponding to the six points are presented in front and side views Fig. 5**D**. Moreover, the normalized height, $H_D/H_{D0}$, and the dihedral angles, $\varphi_i$ ($i$=1, 2, 3, …, 12), of the structure, are respectively shown in Fig. 5**E** and **F**. Finally, the total energy of the facets and that of the creases are drawn in Fig. 5**G**. In the calculation of crease energy, a narrow strip of 0.6mm on each side of the crease is included so that the width is 1.2mm which is equal to that of the physical specimen. Apart from the global behavior of the structure, the diagonal lengths, energies, as well as the von Mises stress and equivalent plastic strain (PEEQ) contours of the local square, rectangular, and trapezoidal facets are presented in Fig. 6 to depict their deformation evolution in detail.

It can be seen from Fig. 5**E** that during the unfolding process, while the diagonal length monotonically increases, the structure height first slightly drops, and then rises quickly followed by another slow drop. Consequently, the deformation process of the structure can be divided into three stages: a tightening stage (configurations I-II), an unlocking stage (configurations II-V), and a flattening stage (configurations V-VI). First, consider the tightening stage. During this stage, the initially tilted rectangular facets tend to be horizontal, leading to a more compact structure with a slightly reduced height. Theoretically, the fully folded configuration of the structure should have a zero height and a diagonal length of 54.37mm. Due to the natural dihedral angle, nevertheless, the structure has a larger height of 7.80mm combined with a smaller diagonal length of 53.30mm. As a result, when stretched, the structure tends to approach its fully folded configuration, although is not able to completely reach that. At configuration II, the structure reaches its smallest height, or the most locked form that can be achieved for the selected material and loading condition. This explains why the maximum stiffness appears here. Both facet distortions and crease rotations are small at this stage.

Upon further tension, the structure starts to open up, entering the unlocking stage between configuration II and V. As can be seen in Fig. 5**D**, the rectangular and trapezoidal facets gradually tilt at this stage, thereby raising the central square and the height of the structure. From configuration II to III, the crease rotations are found to be non-synchronized, i.e., the dihedral angles formed along the long sides of the rectangular facets, $\varphi_i$ ($i$=3, 6, 9, 12), are almost unchanged whereas all the others increase monotonically, see Fig. 5**F**. To accommodate this non-synchronization, the facets are further distorted to maintain the internal connectivity of the structure. Specifically, the square facet is squeezed by the neighboring facets and bulges out-of-plane to form a dome-like shape, which is manifested by the shrinkage in diagonal length in Fig. 6**A**. The corner areas undergo the largest deformation and develop plasticity shown in the PEEQ contour Fig. 6**G**. Meanwhile, the rectangular and trapezoidal facets are also minorly compressed along the diagonal lengths as shown in Fig. 6**B** and **C**, but remain elastic in most of the area. The large facet distortions lead to a sharp rise in the reaction force, which reaches its initial peak force at configuration III. From configuration III to IV, creases 3, 6, 9, 12 start to open up along with the others, which in turn mitigate the required facet distortions. As a result, the deformation of the facets continues to develop, but at a gradually reduced rate, which can be concluded by comparing the energy curves of the three facets between II-III and III-IV in Fig. 6**D-F**. This leads to a reduced reaction force and consequently a negative stiffness. When the unit passes configuration III, large plastic regions start to appear in the three types of facets, especially the rectangular one, see the PEEQ contours of in Fig. 6**G-I**. This is echoed in the plastic energy of the facets in Fig. 6**D-F**. The development of plasticity slowers the energy development of the structure, leading to the transition point IV on the energy curve in Fig. 5**A**. Further stretching the structure to configuration V, the diagonal length of the rectangular facet reaches its minimum, indicating that it reaches its most distorted, also most tilted shape. This marks the closure of the unlocking stage, characterized by the largest structure height.

After configuration V, the structure enters the final flattening stage. At this stage, the tilted rectangular and trapezoidal facets rotate along the creases to level again, which causes a further unfolding of the structure accompanied by a reduction in height. The energy curves in Fig. 6**D-F** indicate that all the facets tend to recover by partially releasing their elastic energy. The rectangular facet which is mostly deformed releases more than half of its elastic energy, whereas its plastic energy keeps steady. All the creases, on the other hand, continuously open up till the end as shown in Fig. 5**F**, resulting in a roughly linearly increasing crease energy curve in Fig. 5**G**. Since the reaction force is mainly used to overcome the stiffness of the creases, a relatively low force in comparison with the previous two stages is generated until when the structure is nearly flat at the end.

Overall, the total crease energy in Fig. 5**G** tends to increase approximately during tension, whereas the total facet energy shows a local peak associated with the unlocking process. Moreover, the crease energy is much larger than the facet energy in the end. This result indicates that facet deformation is mainly responsible for the high initial peak, while crease rotation is still the main source of structure energy absorption.

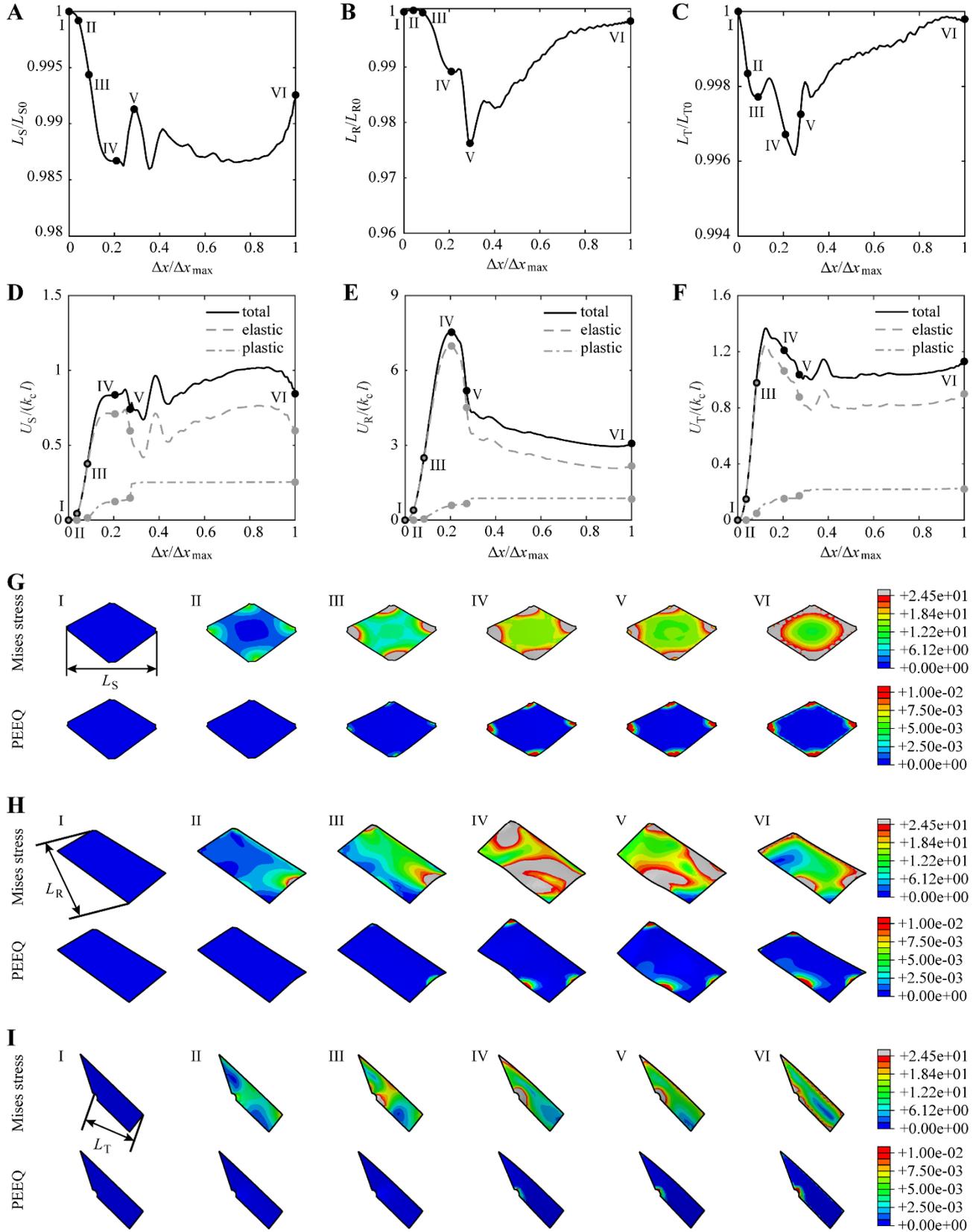

**Fig. 6.** Illustration of the deformed square, rectangular, and trapezoidal facets. (**A-C**) Normalized diagonal length of the square facet, $L_S/L_{S0}$, that of the rectangular facet, $L_R/L_{R0}$, and that of the trapezoidal facet, $L_T/L_{T0}$. (**D-F**) Normalized energy of the square facet, $U_S/(k_c l)$, that of the rectangular facet, $U_R/(k_c l)$, and that of the trapezoidal facet, $U_T/(k_c l)$, versus normalized displacement, $\Delta x/\Delta x_{\max}$. (**G-I**) The Mises stress and PEEQ contours of the square, rectangular, and trapezoidal facets.

## 4. Mechanical properties programmability

### 4.1. Development of empirical model

From the viewpoint of programmability, it is crucial to develop a mathematical model to predict the mechanical properties of the type 1 square-twist structure. However, it has been shown in Section 3 that very complicated deformation modes are generated on the facets of the structure during the unfolding process. As a result, it is difficult to build an elegant theoretical model for the structure by previous methods such as adding virtual creases to create an equivalent rigid structure. Instead, we develop an empirical model based on the numerical results to provide a practical prediction approach. Here, three key mechanical properties are focused on, i.e. the energy at the end of deformation corresponding to configuration VI, the initial peak force corresponding to configuration III, and the maximum stiffness corresponding to configuration II in Fig. 5**A-C**. We start by analyzing the energy of the structure which is the summation of the total crease energy and the total facet energy. The energy of a crease is dependent on its torsional stiffness per unit length, length, as well as the amount of rotation. As a result, we assume that the total crease energy to be in the form of

$$U_c = \sum_{i=1}^{12} u_{ci} \cdot k_{ci} \cdot L_{ci} \cdot f_{ci}(\theta) \tag{1}$$

where $u_{ci}$, $k_{ci}$, $L_{ci}$, and $f_{ci}(\theta)$ are respectively a constant energy coefficient, torsional stiffness per unit length, crease length, and rotation function with a variable $\theta$.

Since the structure has a four-fold rotational symmetry and all the creases have identical stiffness, all the 12 creases fall into three groups with length $l$ represented by $\varphi_i$ ($i$=1, 4, 7, 10), $a$ represented by $\varphi_i$ ($i$=2, 5, 8, 11), and $l \cdot \cos\alpha + a$ represented by $\varphi_i$ ($i$=3, 6, 9, 12). Thus we can rewrite Eq. (1) as

$$U_c = 4k_c [u_{c1} \cdot l \cdot f_{c1}(\theta) + u_{c2} \cdot a \cdot f_{c2}(\theta) + u_{c3} \cdot (l \cdot \cos\alpha + a) \cdot f_{c3}(\theta)] \tag{2}$$

Notice that there are three deformation functions in the equation because the three groups of creases do not open up at the same rate, which can be seen in Fig. 5**F**.

Regarding the facets, it has been shown from the numerical simulation that both elastic and plastic regions could develop in the facets during loading. Therefore, we first assume the plastic regions of the square, rectangular, and trapezoidal facets, which depend on the geometric and material parameters, as follows

$$S_{S\text{-}p} = \gamma_{S1} \cdot l^2 \cdot \tan\alpha + \gamma_{S2} \cdot a^2 + \gamma_{S3} \cdot k_c \cdot l^2 \tag{3}$$

$$S_{R\text{-}p} = \gamma_{R1} \cdot l^2 \cdot \tan\alpha + \gamma_{R2} \cdot a^2 + \gamma_{R3} \cdot k_c \cdot l^2 \tag{4}$$

$$S_{T\text{-}p} = \gamma_{T1} \cdot l^2 \cdot \tan\alpha + \gamma_{T2} \cdot a^2 + \gamma_{T3} \cdot k_c \cdot l^2 \tag{5}$$

in which $\gamma_{S1}$, $\gamma_{S2}$, $\gamma_{S3}$, $\gamma_{R1}$, $\gamma_{R2}$, $\gamma_{R3}$, $\gamma_{T1}$, $\gamma_{T2}$, and $\gamma_{T3}$ are constant coefficients.

Then the elastic regions of the three types of facets are

$$S_{S\text{-}e} = l^2 - \gamma_{S1} \cdot l^2 \cdot \tan\alpha - \gamma_{S2} \cdot a^2 - \gamma_{S3} \cdot k_c \cdot l^2 \tag{6}$$

$$S_{R\text{-}e} = a \cdot (l \cdot \cos\alpha + a) - \gamma_{R1} \cdot l^2 \cdot \tan\alpha - \gamma_{R2} \cdot a^2 - \gamma_{R3} \cdot k_c \cdot l^2 \tag{7}$$

$$S_{T\text{-}e} = l \cdot \sin\alpha \cdot (l \cdot \cos\alpha + 2a) - \gamma_{T1} \cdot l^2 \cdot \tan\alpha - \gamma_{T2} \cdot a^2 - \gamma_{T3} \cdot k_c \cdot l^2 \tag{8}$$

With the plastic and elastic regions, the total facet energy is assumed to be calculated by the following equation

$$\begin{aligned} U_f = &\, u_{S\text{-}e} \cdot \left(l^2 - \gamma_{S1} \cdot l^2 \cdot \tan\alpha - \gamma_{S2} \cdot a^2 - \gamma_{S3} \cdot k_c \cdot l^2\right) \cdot f_S(\theta) \\ &+ u_{S\text{-}p} \cdot \left(\gamma_{S1} \cdot l^2 \cdot \tan\alpha + \gamma_{S2} \cdot a^2 + \gamma_{S3} \cdot k_c \cdot l^2\right) \cdot f_S(\theta) \\ &+ 4u_{R\text{-}e} \cdot \left[a \cdot (l \cdot \cos\alpha + a) - \gamma_{R1} \cdot l^2 \cdot \tan\alpha - \gamma_{R2} \cdot a^2 - \gamma_{R3} \cdot k_c \cdot l^2\right] \cdot f_R(\theta) \\ &+ 4u_{R\text{-}p} \cdot \left(\gamma_{R1} \cdot l^2 \cdot \tan\alpha + \gamma_{R2} \cdot a^2 + \gamma_{R3} \cdot k_c \cdot l^2\right) \cdot f_R(\theta) \\ &+ 4u_{T\text{-}e} \cdot \left[l \cdot \sin\alpha \cdot (l \cdot \cos\alpha + 2a) - \gamma_{T1} \cdot l^2 \cdot \tan\alpha - \gamma_{T2} \cdot a^2 - \gamma_{T3} \cdot k_c \cdot l^2\right] \cdot f_T(\theta) \\ &+ 4u_{T\text{-}p} \cdot \left(\gamma_{T1} \cdot l^2 \cdot \tan\alpha + \gamma_{T2} \cdot a^2 + \gamma_{T3} \cdot k_c \cdot l^2\right) \cdot f_T(\theta) \end{aligned} \tag{9}$$

where $u_{S\text{-}e}$, $u_{R\text{-}e}$, $u_{T\text{-}e}$, and $u_{S\text{-}p}$, $u_{R\text{-}p}$, $u_{T\text{-}p}$ are respectively the elastic and plastic energy coefficients of the square, rectangular, and trapezoidal facets, and $f_S(\theta)$, $f_R(\theta)$, $f_T(\theta)$ are the deformation functions.

The diagonal length of the structure is assumed to be related to the crease lengths and rotations through the following equation

$$D = \sqrt{2}[d_1 \cdot l \cdot \sin\alpha \cdot f_{d1}(\theta) + d_2 \cdot a \cdot f_{d2}(\theta) + d_3 \cdot (l \cdot \cos\alpha + a) \cdot f_{d3}(\theta)] \tag{10}$$

where $d_1$, $d_2$, and $d_3$ are length coefficients, $f_{d1}(\theta)$, $f_{d2}(\theta)$, and $f_{d3}(\theta)$ are deformation functions of creases 1, 2, and 3, respectively. Notice that the rotation of crease 2 does not affect the diagonal length, and therefore $f_{d2}(\theta)=1$. Consequently, Eq. (10) can be simplified to

$$D = \sqrt{2}[d_1 \cdot l \cdot \sin\alpha \cdot f_{d1}(\theta) + d_2 \cdot a + d_3 \cdot (l \cdot \cos\alpha + a) \cdot f_{d3}(\theta)] \tag{11}$$

With all the above equations, the energy, force, and stiffness of the structure can be obtained as

$$\begin{aligned}
U =\ & u_{S\text{-}e} \cdot (l^2 - \gamma_{S1} \cdot l^2 \cdot \tan\alpha - \gamma_{S2} \cdot a^2 - \gamma_{S3} \cdot k_c \cdot l^2) \cdot f_S(\theta) \\
& + u_{S\text{-}p} \cdot (\gamma_{S1} \cdot l^2 \cdot \tan\alpha + \gamma_{S2} \cdot a^2 + \gamma_{S3} \cdot k_c \cdot l^2) \cdot f_S(\theta) \\
& + 4u_{R\text{-}e} \cdot [a \cdot (l \cdot \cos\alpha + a) - \gamma_{R1} \cdot l^2 \cdot \tan\alpha - \gamma_{R2} \cdot a^2 - \gamma_{R3} \cdot k_c \cdot l^2] \cdot f_R(\theta) \\
& + 4u_{R\text{-}p} \cdot (\gamma_{R1} \cdot l^2 \cdot \tan\alpha + \gamma_{R2} \cdot a^2 + \gamma_{R3} \cdot k_c \cdot l^2) \cdot f_R(\theta) \\
& + 4u_{T\text{-}e} \cdot [l \cdot \sin\alpha \cdot (l \cdot \cos\alpha + 2a) - \gamma_{T1} \cdot l^2 \cdot \tan\alpha - \gamma_{T2} \cdot a^2 - \gamma_{T3} \cdot k_c \cdot l^2] \cdot f_T(\theta) \\
& + 4u_{T\text{-}p} \cdot (\gamma_{T1} \cdot l^2 \cdot \tan\alpha + \gamma_{T2} \cdot a^2 + \gamma_{T3} \cdot k_c \cdot l^2) \cdot f_T(\theta) \\
& + 4u_{c1} \cdot k_c \cdot a \cdot f_{c1}(\theta) + 4u_{c2} \cdot k_c \cdot l \cdot f_{c2}(\theta) + 4u_{c3} \cdot k_c \cdot (l \cdot \cos\alpha + a) \cdot f_{c3}(\theta)
\end{aligned} \tag{12}$$

$$F = \frac{dU/d\theta}{dD/d\theta} = \frac{\begin{bmatrix} u_{S\text{-}e} \cdot (l^2 - \gamma_{S1} \cdot l^2 \cdot \tan\alpha - \gamma_{S2} \cdot a^2 - \gamma_{S3} \cdot k_c \cdot l^2) \cdot f'_S(\theta) \\ + u_{S\text{-}p} \cdot (\gamma_{S1} \cdot l^2 \cdot \tan\alpha + \gamma_{S2} \cdot a^2 + \gamma_{S3} \cdot k_c \cdot l^2) \cdot f'_S(\theta) \\ + 4u_{R\text{-}e} \cdot [a \cdot (l \cdot \cos\alpha + a) - \gamma_{R1} \cdot l^2 \cdot \tan\alpha - \gamma_{R2} \cdot a^2 - \gamma_{R3} \cdot k_c \cdot l^2] \cdot f'_R(\theta) \\ + 4u_{R\text{-}p} \cdot (\gamma_{R1} \cdot l^2 \cdot \tan\alpha + \gamma_{R2} \cdot a^2 + \gamma_{R3} \cdot k_c \cdot l^2) \cdot f'_R(\theta) \\ + 4u_{T\text{-}e} \cdot [l \cdot \sin\alpha \cdot (l \cdot \cos\alpha + 2a) - \gamma_{T1} \cdot l^2 \cdot \tan\alpha - \gamma_{T2} \cdot a^2 - \gamma_{T3} \cdot k_c \cdot l^2] \cdot f'_T(\theta) \\ + 4u_{T\text{-}p} \cdot (\gamma_{T1} \cdot l^2 \cdot \tan\alpha + \gamma_{T2} \cdot a^2 + \gamma_{T3} \cdot k_c \cdot l^2) \cdot f'_T(\theta) \\ + 4u_{c1} \cdot k_c \cdot a \cdot f'_{c1}(\theta) + 4u_{c2} \cdot k_c \cdot l \cdot f'_{c2}(\theta) + 4u_{c3} \cdot k_c \cdot (l \cdot \cos\alpha + a) \cdot f'_{c3}(\theta) \end{bmatrix}}{d_1 \cdot l \cdot \sin\alpha \cdot f'_{d1}(\theta) + d_3 \cdot (l \cdot \cos\alpha + a) \cdot f'_{d3}(\theta)} \tag{13}$$

$$K = \frac{dF/d\theta}{dD/d\theta} = \frac{\begin{bmatrix} Q_1 \cdot (d_1 \cdot l \cdot \sin\alpha \cdot f'_{d1}(\theta) + d_3 \cdot (l \cdot \cos\alpha + a) \cdot f'_{d3}(\theta)) \\ - Q_2 \cdot (d'_1 \cdot l \cdot \sin\alpha \cdot f''_{d1}(\theta) + d'_3 \cdot (l \cdot \cos\alpha + a) \cdot f''_{d3}(\theta)) \end{bmatrix}}{[d_1 \cdot l \cdot \sin\alpha \cdot f'_{d1}(\theta) + d_3 \cdot (l \cdot \cos\alpha + a) \cdot f'_{d3}(\theta)]^3} \tag{14}$$

where

$$\begin{aligned}
Q_1 =\ & u_{S\text{-}e} \cdot \left(l^2 - \gamma_{S1} \cdot l^2 \cdot \tan\alpha - \gamma_{S2} \cdot a^2 - \gamma_{S3} \cdot k_c \cdot l^2\right) \cdot f_S''(\theta) \\
& + u_{S\text{-}p} \cdot \left(\gamma_{S1} \cdot l^2 \cdot \tan\alpha + \gamma_{S2} \cdot a^2 + \gamma_{S3} \cdot k_c \cdot l^2\right) \cdot f_S''(\theta) \\
& + 4u_{R\text{-}e} \cdot \left[a \cdot (l \cdot \cos\alpha + a) - \gamma_{R1} \cdot l^2 \cdot \tan\alpha - \gamma_{R2} \cdot a^2 - \gamma_{R3} \cdot k_c \cdot l^2\right] \cdot f_R''(\theta) \\
& + 4u_{R\text{-}p} \cdot \left(\gamma_{R1} \cdot l^2 \cdot \tan\alpha + \gamma_{R2} \cdot a^2 + \gamma_{R3} \cdot k_c \cdot l^2\right) \cdot f_R''(\theta) \\
& + 4u_{T\text{-}e} \cdot \left[l \cdot \sin\alpha \cdot (l \cdot \cos\alpha + 2a) - \gamma_{T1} \cdot l^2 \cdot \tan\alpha - \gamma_{T2} \cdot a^2 - \gamma_{T3} \cdot k_c \cdot l^2\right] \cdot f_T''(\theta) \\
& + 4u_{T\text{-}p} \cdot \left(\gamma_{T1} \cdot l^2 \cdot \tan\alpha + \gamma_{T2} \cdot a^2 + \gamma_{T3} \cdot k_c \cdot l^2\right) \cdot f_T''(\theta) \\
& + 4u_{c1} \cdot k_c \cdot a \cdot f_{c1}''(\theta) + 4u_{c2} \cdot k_c \cdot l \cdot f_{c2}''(\theta) + 4u_{c3} \cdot k_c \cdot (l \cdot \cos\alpha + a) \cdot f_{c3}''(\theta)
\end{aligned} \tag{15}$$

$$\begin{aligned}
Q_2 =\ & u_{S\text{-}e} \cdot \left(l^2 - \gamma_{S1} \cdot l^2 \cdot \tan\alpha - \gamma_{S2} \cdot a^2 - \gamma_{S3} \cdot k_c \cdot l^2\right) \cdot f_S'(\theta) \\
& + u_{S\text{-}p} \cdot \left(\gamma_{S1} \cdot l^2 \cdot \tan\alpha + \gamma_{S2} \cdot a^2 + \gamma_{S3} \cdot k_c \cdot l^2\right) \cdot f_S'(\theta) \\
& + 4u_{R\text{-}e} \cdot \left[a \cdot (l \cdot \cos\alpha + a) - \gamma_{R1} \cdot l^2 \cdot \tan\alpha - \gamma_{R2} \cdot a^2 - \gamma_{R3} \cdot k_c \cdot l^2\right] \cdot f_R'(\theta) \\
& + 4u_{R\text{-}p} \cdot \left(\gamma_{R1} \cdot l^2 \cdot \tan\alpha + \gamma_{R2} \cdot a^2 + \gamma_{R3} \cdot k_c \cdot l^2\right) \cdot f_R'(\theta) \\
& + 4u_{T\text{-}e} \cdot \left[l \cdot \sin\alpha \cdot (l \cdot \cos\alpha + 2a) - \gamma_{T1} \cdot l^2 \cdot \tan\alpha - \gamma_{T2} \cdot a^2 - \gamma_{T3} \cdot k_c \cdot l^2\right] \cdot f_T'(\theta) \\
& + 4u_{T\text{-}p} \cdot \left(\gamma_{T1} \cdot l^2 \cdot \tan\alpha + \gamma_{T2} \cdot a^2 + \gamma_{T3} \cdot k_c \cdot l^2\right) \cdot f_T'(\theta) \\
& + 4u_{c1} \cdot k_c \cdot a \cdot f_{c1}'(\theta) + 4u_{c2} \cdot k_c \cdot l \cdot f_{c2}'(\theta) + 4u_{c3} \cdot k_c \cdot (l \cdot \cos\alpha + a) \cdot f_{c3}'(\theta)
\end{aligned} \tag{16}$$

Assuming the variable $\theta$ corresponding to the energy at the end of loading, initial peak force, and maximum stiffness is $\theta_U$, $\theta_F$, and $\theta_K$, respectively, the three properties can be calculated from Eq. (12)-(14),

$$\begin{aligned}
U =\ & u_{U1} \cdot \left(l^2 - \gamma_{S1} \cdot l^2 \cdot \tan\alpha - \gamma_{S2} \cdot a^2 - \gamma_{S3} \cdot k_c \cdot l^2\right) + u_{U2} \cdot \left(\gamma_{S1} \cdot l^2 \cdot \tan\alpha + \gamma_{S2} \cdot a^2 + \gamma_{S3} \cdot k_c \cdot l^2\right) \\
& + u_{U3} \cdot \left[a \cdot (l \cdot \cos\alpha + a) - \gamma_{R1} \cdot l^2 \cdot \tan\alpha - \gamma_{R2} \cdot a^2 - \gamma_{R3} \cdot k_c \cdot l^2\right] \\
& + u_{U4} \cdot \left(\gamma_{R1} \cdot l^2 \cdot \tan\alpha + \gamma_{R2} \cdot a^2 + \gamma_{R3} \cdot k_c \cdot l^2\right) \\
& + u_{U5} \cdot \left[l \cdot \sin\alpha \cdot (l \cdot \cos\alpha + 2a) - \gamma_{T1} \cdot l^2 \cdot \tan\alpha - \gamma_{T2} \cdot a^2 - \gamma_{T3} \cdot k_c \cdot l^2\right] \\
& + u_{U6} \cdot \left(\gamma_{T1} \cdot l^2 \cdot \tan\alpha + \gamma_{T2} \cdot a^2 + \gamma_{T3} \cdot k_c \cdot l^2\right) \\
& + u_{U7} \cdot k_c \cdot a + u_{U8} \cdot k_c \cdot l + u_{U9} \cdot k_c \cdot (l \cdot \cos\alpha + a)
\end{aligned} \tag{17}$$

$$F_{\max} = \frac{\begin{bmatrix} u_{F1} \cdot \left(l^2 - \gamma_{S1} \cdot l^2 \cdot \tan\alpha - \gamma_{S2} \cdot a^2 - \gamma_{S3} \cdot k_c \cdot l^2\right) \\ + u_{F2} \cdot \left(\gamma_{S1} \cdot l^2 \cdot \tan\alpha + \gamma_{S2} \cdot a^2 + \gamma_{S3} \cdot k_c \cdot l^2\right) \\ + u_{F3} \cdot \left(a \cdot (l \cdot \cos\alpha + a) - \gamma_{R1} \cdot l^2 \cdot \tan\alpha - \gamma_{R2} \cdot a^2 - \gamma_{R3} \cdot k_c \cdot l^2\right) \\ + u_{F4} \cdot \left(\gamma_{R1} \cdot l^2 \cdot \tan\alpha + \gamma_{R2} \cdot a^2 + \gamma_{R3} \cdot k_c \cdot l^2\right) \\ + u_{F5} \cdot \left(l \cdot \sin\alpha \cdot (l \cdot \cos\alpha + 2a) - \gamma_{T1} \cdot l^2 \cdot \tan\alpha - \gamma_{T2} \cdot a^2 - \gamma_{T3} \cdot k_c \cdot l^2\right) \\ + u_{F6} \cdot \left(\gamma_{T1} \cdot l^2 \cdot \tan\alpha + \gamma_{T2} \cdot a^2 + \gamma_{T3} \cdot k_c \cdot l^2\right) \\ + u_{F7} \cdot k_c \cdot a + u_{F8} \cdot k_c \cdot l + u_{F9} \cdot k_c \cdot (l \cdot \cos\alpha + a) \end{bmatrix}}{d_{F1} \cdot l \cdot \sin\alpha + d_{F3} \cdot (l \cdot \cos\alpha + a)} \tag{18}$$

$$K_{\max} = \frac{Q_1 \cdot [d_{K1} \cdot l \cdot \sin\alpha + d_{K3} \cdot (l \cdot \cos\alpha + a)] - Q_2 \cdot [d'_{K1} \cdot l \cdot \sin\alpha + d'_{K3} \cdot (l \cdot \cos\alpha + a)]}{[d_{K1} \cdot l \cdot \sin\alpha + d_{K3} \cdot (l \cdot \cos\alpha + a)]^3} \tag{19}$$

where

$$\begin{aligned} Q_1 &= u_{K1} \cdot l^2 + u_{K2} \cdot a \cdot (l \cdot \cos\alpha + a) + u_{K3} \cdot l \cdot \sin\alpha \cdot (l \cdot \cos\alpha + 2a) \\ &+ u_{K4} \cdot k_c \cdot a + u_{K5} \cdot k_c \cdot l + u_{K6} \cdot k_c \cdot (l \cdot \cos\alpha + a) \end{aligned} \tag{20}$$

$$\begin{aligned} Q_2 &= u_{K7} \cdot l^2 + u_{K8} \cdot a \cdot (l \cdot \cos\alpha + a) + u_{K9} \cdot l \cdot \sin\alpha \cdot (l \cdot \cos\alpha + 2a) \\ &+ u_{K10} \cdot k_c \cdot a + u_{K11} \cdot k_c \cdot l + u_{K12} \cdot k_c \cdot (l \cdot \cos\alpha + a) \end{aligned} \tag{21}$$

$$\begin{aligned} &u_{U1} = u_{S\text{-}e} \cdot f_S(\theta)|_{\theta_U}, u_{U2} = u_{S\text{-}p} \cdot f_S(\theta)|_{\theta_U}, u_{U3} = 4u_{R\text{-}e} \cdot f_R(\theta)|_{\theta_U}, \\ &u_{U4} = 4u_{R\text{-}p} \cdot f_R(\theta)|_{\theta_U}, u_{U5} = 4u_{T\text{-}e} \cdot f_T(\theta)|_{\theta_U}, u_{U6} = 4u_{T\text{-}p} \cdot f_T(\theta)|_{\theta_U}, \\ &u_{U7} = 4u_{c1} \cdot f_{c1}(\theta)|_{\theta_U}, u_{U8} = 4u_{c2} \cdot f_{c2}(\theta)|_{\theta_U}, u_{U9} = 4u_{c3} \cdot f_{c3}(\theta)|_{\theta_U}; \end{aligned} \tag{22}$$

$$\begin{aligned} &u_{F1} = u_{S\text{-}e} \cdot f'_S(\theta)|_{\theta_F}, u_{F2} = u_{S\text{-}p} \cdot f'_S(\theta)|_{\theta_F}, u_{F3} = 4u_{R\text{-}e} \cdot f'_R(\theta)|_{\theta_F}, \\ &u_{F4} = 4u_{R\text{-}p} \cdot f'_R(\theta)|_{\theta_F}, u_{F5} = 4u_{T\text{-}e} \cdot f'_T(\theta)|_{\theta_F}, u_{F6} = 4u_{T\text{-}p} \cdot f'_T(\theta)|_{\theta_F}, \\ &u_{F7} = 4u_{c1} \cdot f'_{c1}(\theta)|_{\theta_F}, u_{F8} = 4u_{c2} \cdot f'_{c2}(\theta)|_{\theta_F}, u_{F9} = 4u_{c3} \cdot f'_{c3}(\theta)|_{\theta_F}, \\ &d_{F1} = d_1 \cdot f'_{d1}(\theta)|_{\theta_F}, d_{F3} = d_3 \cdot f'_{d3}(\theta)|_{\theta_F}; \end{aligned} \tag{23}$$

$$\begin{aligned} &u_{K1} = u_{S\text{-}e} \cdot f''_S(\theta)|_{\theta_K}, u_{K2} = 4u_{R\text{-}e} \cdot f''_R(\theta)|_{\theta_K}, u_{K3} = 4u_{T\text{-}e} \cdot f''_T(\theta)|_{\theta_K}, \\ &u_{K4} = 4u_{c1} \cdot f''_{c1}(\theta)|_{\theta_K}, u_{K5} = 4u_{c2} \cdot f''_{c2}(\theta)|_{\theta_K}, u_{K6} = 4u_{c3} \cdot f''_{c3}(\theta)|_{\theta_K}, \\ &u_{K7} = u_{S\text{-}e} \cdot f'_S(\theta)|_{\theta_K}, u_{K8} = 4u_{R\text{-}e} \cdot f'_R(\theta)|_{\theta_K}, u_{K9} = 4u_{T\text{-}e} \cdot f'_T(\theta)|_{\theta_K}, \\ &u_{K10} = 4u_{c1} \cdot f'_{c1}(\theta)|_{\theta_K}, u_{K11} = 4u_{c2} \cdot f'_{c2}(\theta)|_{\theta_K}, u_{K12} = 4u_{c3} \cdot f'_{c3}(\theta)|_{\theta_K}, \\ &d_{K1} = d_1 \cdot f'_{d1}(\theta)|_{\theta_K}, d_{K3} = d_3 \cdot f'_{d3}(\theta)|_{\theta_K}, d'_{K1} = d'_1 \cdot f''_{d1}(\theta)|_{\theta_K}, d'_{K3} = d'_3 \cdot f''_{d3}(\theta)|_{\theta_K}. \end{aligned} \tag{24}$$

Notice that in Eq. (19) for the maximum stiffness, the terms associated with the plastic regions are ignored because the PEEQ contours in Fig. 6**G-I** indicate that all the facets remain elastic at configuration II where the maximum stiffness is achieved.

To determine the unknown coefficients in Eq. (17)-(21), a series of 20 numerical models with varying side length, *a*, twist angle, *α*, and crease torsional stiffness, $k_c$, as listed in Table 1 were built and analyzed. The side length, *l*, was fixed to 16.25mm for all the models, and the sheet thickness and material parameters were the same as that in Section 2. Based on the results also listed in Table 1, the coefficients can be obtained as follows using the nonlinear regression.

$$u_{U1} = -0.92, u_{U2} = 7.1\times10^{-5}, u_{U3} = 0.21, u_{U4} = -17.66, u_{U5} = -0.39, u_{U6} = 11.69,$$
$$u_{U7} = 6.73, u_{U8} = 5.57\times10^{-2}, u_{U9} = 8.54, \gamma_{S1} = 0.32, \gamma_{S2} = 6\times10^{-6}, \gamma_{S3} = 0.27,$$
$$\gamma_{R1} = 4.2\times10^{-6}, \gamma_{R2} = 9.26\times10^{-3}, \gamma_{R3} = 0.12, \gamma_{T1} = 0.18, \gamma_{T2} = 8.1\times10^{-6}, \gamma_{T3} = 0.17; \quad (25)$$

$$u_{F1} = -2.18, u_{F2} = 12.7, u_{F3} = 0.29, u_{F4} = 13.49, u_{F5} = 1.44, u_{F6} = 9.71,$$
$$u_{F7} = 7.6, u_{F8} = 8.74\times10^{-2}, u_{F9} = 2.02\times10^{-2}, d_{F1} = -1.52, d_{F3} = 0.9,$$
$$\gamma_{S1} = 3.3\times10^{-2}, \gamma_{S2} = 6.5\times10^{-3}, \gamma_{S3} = 9\times10^{-9}, \gamma_{R1} = 6.5\times10^{-2}, \gamma_{R2} = 4.2\times10^{-3}, \quad (26)$$
$$\gamma_{R3} = 1.6\times10^{-9}, \gamma_{T1} = 8.3\times10^{-2}, \gamma_{T2} = 8.1\times10^{-3}, \gamma_{T3} = 5.8\times10^{-10};$$

$$u_{K1} = -7.14, u_{K2} = 3.44, u_{K3} = 1.93, u_{K4} = 9.67, u_{K5} = 5.72\times10^{-4}, u_{K6} = 1.19,$$
$$u_{K7} = -2.57, u_{K8} = -2.72, u_{K9} = 17.57, u_{K10} = 0.016, u_{K11} = 4.51, u_{K12} = 2.04\times10^{-3}, \quad (27)$$
$$d_{K1} = -0.11, d_{K3} = 0.25, d'_{K1} = -0.56, d'_{K3} = 0.11.$$

Substituting Eq. (25)-(27) to (17)-(21), the energy, initial peak force, and maximum stiffness of the structure can be obtained.

The predicted curves of normalized energy, initial peak force, and maximum stiffness of the structure based on the empirical equations (17)-(21) are respectively drawn in Fig. 7**A-C** together with the numerical results (black triangles) and the experimental result (red circles) in Section 2, from which a good match is obtained for all the three properties. To further validate the accuracy of the empirical equations, another structure with $a$= 24.38mm, $l$=16.25mm, and $\alpha$=35° are fabricated and tested following the same procedure in Section 2. The experimentally obtained energy, initial peak force, and maximum stiffness are 199.73J, 40.28N, and 42.83N·mm$^{-1}$, whereas the predicted values are 202.57J, 54.02N, and 43.73N·mm$^{-1}$. Again the predictions are very close to the corresponding experimental results. Thus we can safely conclude that the empirical equations developed here are capable of predicting the mechanical properties of the type 1 square-twist structure.

**Table 1.** Geometric and material paramters of the numerical models and results.

| Model | $a$ [mm] | $\alpha$ [deg] | $k_c$ [N·rad$^{-1}$] | $U/(k_c l)$ | $F_{max}/k_c$ | $K_{max}l/k_c$ |
|---|---|---|---|---|---|---|
| 1 | 16.25 | 30 | 0.0068 | 202.18 | 4643.79 | 64760.59 |
| 2 | 16.25 | 30 | 0.043 | 69.57 | 750.27 | 10422.74 |
| 3 | 16.25 | 30 | 0.085 | 51.16 | 378.97 | 5252.42 |
| 4 | 16.25 | 30 | 0.11 | 46.8 | 285.42 | 3951.94 |
| 5 | 16.25 | 30 | 0.17 | 41.6 | 192.87 | 2653.91 |
| 6 | 16.25 | 30 | 0.34 | 32.07 | 98.62 | 1330.74 |
| 7 | 16.25 | 30 | 0.51 | 27.26 | 67.15 | 899.22 |
| 8 | 16.25 | 30 | 0.68 | 23.85 | 51.15 | 679.44 |
| 9 | 8.13 | 30 | 0.44 | 23.04 | 53.46 | 995.56 |
| 10 | 12.19 | 30 | 0.44 | 26.08 | 66.2 | 1011.09 |
| 11 | 16.25 | 30 | 0.44 | 28.77 | 77.57 | 1050.77 |
| 12 | 24.38 | 30 | 0.44 | 36.52 | 99.98 | 1223.23 |
| 13 | 32.5 | 30 | 0.44 | 46.34 | 115.33 | 1300.21 |
| 14 | 48.75 | 30 | 0.44 | 63.86 | 142.8 | 1614.16 |
| 15 | 65 | 30 | 0.44 | 72.17 | 145.37 | 1631.23 |
| 16 | 16.25 | 20 | 0.44 | 24.76 | 22.41 | 249.98 |
| 17 | 16.25 | 25 | 0.44 | 27.04 | 41.76 | 544.72 |
| 18 | 16.25 | 35 | 0.44 | 34.72 | 137.75 | 2063.27 |
| 19 | 16.25 | 40 | 0.44 | 50.45 | 201.23 | 3398.61 |
| 20 | 16.25 | 45 | 0.44 | 68.27 | 267.75 | 5055.34 |

**4.2 Properties programmability**

The empirical equations enable us to program the mechanical properties of the structure based on material and geometric parameters. Figure 7**A** shows the results with fixed geometric parameters $a/l=1$ and $\alpha=30°$, and varying ratio of facet bending stiffness to crease rotational stiffness $k_f/k_c$, in which $k_f$ is experimentally determined to be 0.70 N·rad$^{-1}$ based on the method in reference [33]. It can be seen that when $k_f/k_c$ is relatively small, the energy increases at a low rate with $k_f/k_c$, which implies that the contribution from crease rotations is dominant in the energy of the structure. As $k_f/k_c$ becomes larger, the role played by facet distortions is more prominent, and consequently the energy rises at a higher rate. The initial peak force and maximum stiffness, on the other hand, increase nearly linearly with $k_f/k_c$ in the entire range, indicating that these two properties are dominated by the facets of the structure.

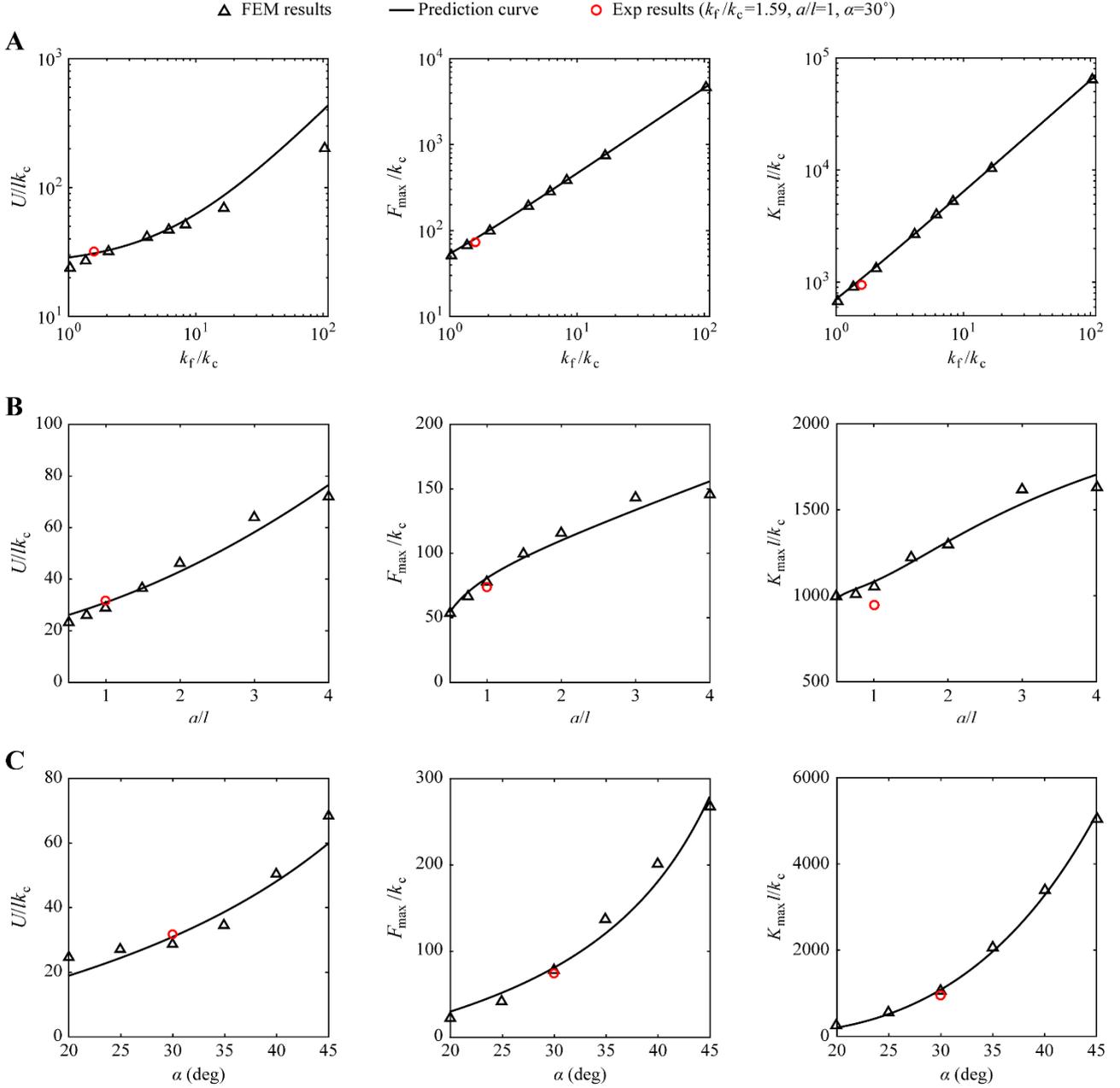

**Fig. 7.** Predicted normalized energy, $U/(k_c l)$, initial peak force, $F_{max}/k_c$, and maximum stiffness $K_{max} l/k_c$, of the type 1 square-twist structure with (**A**) normalized crease stiffness $k_f/k_c$ from 1 to 110, (**B**) side length ratio $a/l$ from 0.5 to 4, and (**C**) twist angle $\alpha$ from 20° to 45°. The experimental and numerical results are also shown in the figure.

Furthermore, if we keep the crease stiffness ratio $k_f/k_c=1.59$ and the twist angle $\alpha=30°$, and vary the side length ratio $a/l$ from 0.5 to 4, we can find out from Fig. 7**B** that the energy, initial peak force, and maximum stiffness increase with $a/l$ since it enlarges the rectangular and trapezoidal facets while keeping the size of the square constant. Finally, the results of the models with identical $k_f/k_c=1.59$ and

*a*/*l*=1, and different twist angle $\alpha$ ranging from 20° to 45°, are shown in Fig. 7**C**. Increasing $\alpha$ raises all three mechanical properties especially the maximum stiffness since a larger twist angle leads to a more twisted and consequently stiffer structure. When $\alpha$ is less than 20°, the structure behaves like a rigid origami structure with no obvious initial peak force but a long plateau force as in the case of the rigid type 3 square-twist structure [34].

## 5. Conclusion

To conclude, in this paper we have analyzed the deformation characteristics and mechanical properties of a rotationally symmetric origami structure, referred to as the type 1 square-twist structure, experimentally and numerically. The deformation process, divided by the tightening stage, unlocking stage, and flattening stage, has been analyzed in detail, together with the key features in the energy, force, and stiffness curves. Based on the deformation analysis of the structure, we have established a series of empirical equations to estimate the energy, initial peak force, and maximum stiffness of the structure, which are validated by quasi-static tension experiments. The empirical model offers an approach to accurately programming the mechanical properties of the non-rigid origami structure by tuning the geometric parameters of the pattern and the base material.

Next, we will study the design principle and property programmability of mechanical metamaterials composed of a single type of square-twist structure or a mixture of them, aiming to develop a series of origami-inspired metamaterials with a wide bandwidth of programmability and tunability.

# Appendix A. Effects of geometric construction methods on the behavior of the type 1 square-twist structure

To investigate the influence of the type of curved surfaces on the behavior of the type 1 square-twist structure, we build and analyze three models in Fig. A.1 besides the one shown in Fig. 3. All the models have the same diagonal length and height as the physical specimen. The model in Fig. A.1**A** has curved rectangular facets bounded by four straight sides. The geometric construction starts with the planar square and rectangular facets based on the diagonal length and height. The loading point is also obtained after this. Subsequently, the trapezoidal facet is introduced to connect with the side of the square facet and the long side of the rectangular facet. Afterward, the planar rectangular facet is removed, and a curved one is defined by the two common sides with the trapezoidal facets and the loading point. Finally, the curved surface of the rectangular facet is generated by the boundary-surface method in Solidworks. For the model in Fig. A.1**B** which has curved trapezoidal facets bounded by four straight sides, the planar square and rectangular facets are first placed based on the diagonal length and height, which determine the two common sides between the rectangular and trapezoidal facets as well as the one between the square and trapezoidal facets. Then, the fourth side of the trapezoidal facet is a straight line between the two endpoints of the common sides with the rectangular facet. Finally, the curved surface of the trapezoidal facet is again generated by the boundary-surface method. For the model in Fig. A.1**C**, the planar square and rectangular facets are placed in the same manner as the model in Fig. A.1**B**. Then, the long common side of the trapezoidal and rectangular facets is cut by a line parallel to the short common side, generating a new curved common side to replace the original straight one. Subsequently, the fourth side of the trapezoidal facet is obtained by connecting the endpoints of the curved side and the short common side with the rectangular facet. Finally, the curved surface of the trapezoidal facet is created using the boundary-surface method. The normalized force versus normalized displacement curves of the four models are shown in Fig. A.1**D**, from which a good agreement is observed. This indicates that the behavior of the structure is not sensitive to the specific type of curved surfaces.

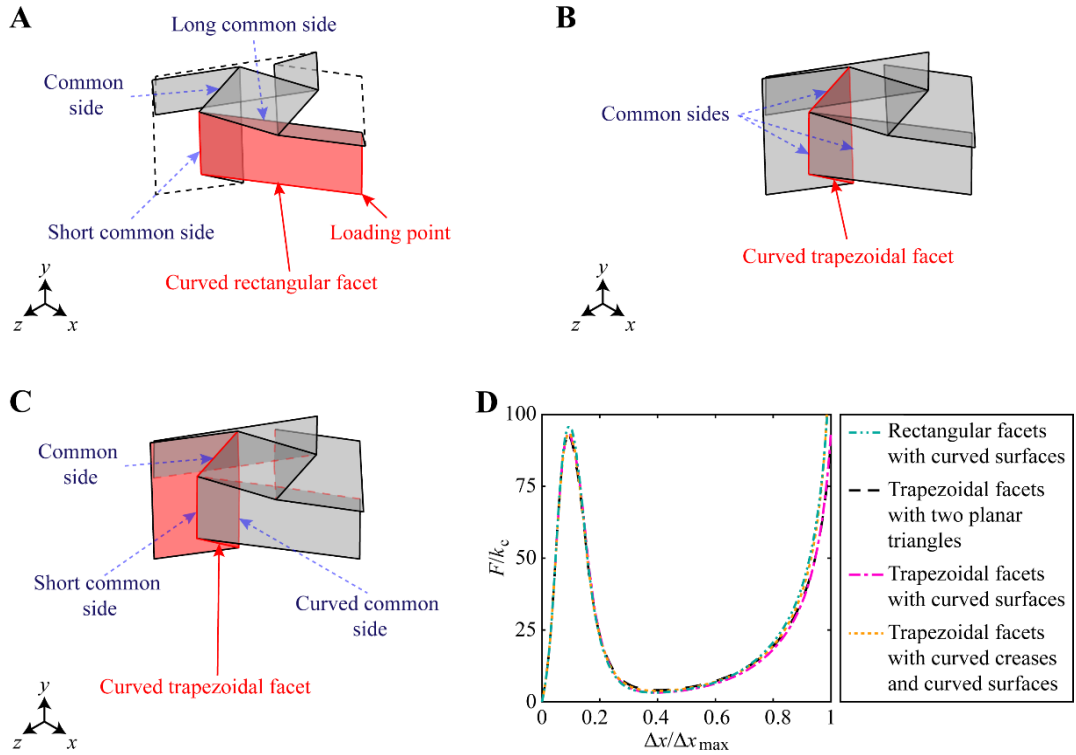

Fig. **A.1.** Effects of geometric construction methods on the behavior of the type 1 square-twist structure, (**A**) replacing the rectangular facets with curved surfaces, (**B**) replacing trapezoidal facets with curved surfaces, and (**C**) replacing trapezoidal facets with curved creases and curved surfaces. (**D**) The normalized force, $F/k_c$, against normalized displacement, $\Delta x/\Delta x_{max}$, of the four models constructed by different methods.

## Acknowledgments

The authors acknowledge the support of the National Natural Science Foundation of China (Projects 52035008, 51825503, 51721003) and the Tencent Foundation (through the XPLORER PRIZE).